\documentclass[journal]{vgtc}                
\ifpdf
  \pdfoutput=1\relax                   
  \usepackage{graphicx}                
  \DeclareGraphicsExtensions{.pdf,.png,.jpg,.jpeg} 
\else
  \usepackage{graphicx}                
  \DeclareGraphicsExtensions{.eps}     
\fi%

\graphicspath{{figures/}{pictures/}{images/}{./}} 

\usepackage{microtype}                 
\PassOptionsToPackage{warn}{textcomp}  
\usepackage{textcomp}                  
\usepackage{mathptmx}                  
\usepackage{times}                     
\usepackage{cite}                      
\usepackage{tabu}                      
\usepackage{booktabs}                  

\onlineid{0}

\vgtccategory{Research}
\vgtcpapertype{please specify}

\PassOptionsToPackage{hyphens}{url}\usepackage{hyperref}
\usepackage{xcolor}
\definecolor{mycolor1}{HTML}{EDF8B1}
\definecolor{mycolor2}{HTML}{C7E9B4}
\definecolor{mycolor3}{HTML}{7FCDBB}
\definecolor{mycolor4}{HTML}{41B6C4}
\definecolor{mycolor5}{HTML}{1D91C0}
\definecolor{mycolor6}{HTML}{225EA8}

\title{Narrative Visualization to Communicate Neurological Diseases}

\author{Sarah Mittenentzwei, Veronika Weiß, Stefanie Schreiber, Laura A. Garrison, \textit{Student Member, IEEE}, \\Stefan Bruckner, \textit{Member, IEEE}, Malte Pfister, Bernhard Preim, and Monique Meuschke 
}
\authorfooter{
\item Sarah Mittenentzwei is with with Otto-von-Guericke University, Magdeburg, Germany, sarah.mittenentzwei@ovgu.de
\item Veronika Weiß is with Freie Universität Berlin, Germany, veronika.weiss@fu-berlin.de
\item Stefanie Schreiber is with the the Department of Neurology, Otto-von-Guericke University and DZNE, Magdeburg, Germany, stefanie.schreiber@med.ovgu.de
\item  Laura A. Garrison is with the University of Bergen, Norway, laura.garrison@uib.no
\item Stefan Bruckner is with the University of Bergen, Norway, stefan.bruckner@uib.no
\item Malte Pfister is with the Department of Neurology, Otto-von-Guericke University, Magdeburg, Germany, malte.pfister@st.ovgu.de
\item Bernhard Preim is with Otto-von-Guericke University, Magdeburg, Germany, bernhard.preim@ovgu.de
\item Monique Meuschke is with Otto-von-Guericke University, Magdeburg, Germany, monique.meuschke@ovgu.de

}

\shortauthortitle{Mittenentzwei \MakeLowercase{\textit{et al.}}: Narrative Visualization to Communicate Neurological Diseases}

\abstract{While narrative visualization has been used successfully in various applications to communicate scientific data in the format of a story to a general audience, the same has not been true for medical data. There are only a few exceptions, such as COVID-19 dashboards that present tabular medical data to non-experts. However, a key component of medical visualization is the interactive analysis of 3D data, such as radiological image volumes or 3D models of anatomical structures, which were rarely included in narrative visualizations so far. 
In this design study, we investigate how neurological disease data can be communicated through narrative visualization techniques to a general audience in an understandable way. In our prototype we designed a narrative visualization explaining cerebral small vessel disease.
Learning about its avoidable risk factors serves to motivate the audience watching the resulting visual data story. Using this example we discuss the adaption of basic narrative components. This includes the conflict and characters of a story, as well as the story's structure and content to address and communicate specific characteristics of medical data. 
Furthermore, we explore the extent to which complex medical relationships need to be simplified to be understandable to a general audience without distorting the underlying data and evidence. In particular, the data needs to be preprocessed for non-experts and appropriate forms of interaction must be found. 
We explore approaches to make the data more personally relatable, such as including a fictional patient. We evaluated our approach in a user study with 40 participants in a web-based implementation of the designed story. We found that the combination of a carefully thought-out storyline with a clear key message, appealing visualizations combined with easy-to-use interactions, and credible references are crucial for creating a narrative visualization about a neurological disease that engages an audience.

} 


\CCScatlist{ 
 \CCScat{K.6.1}{Management of Computing and Information Systems}%
{Project and People Management}{Life Cycle};
 \CCScat{K.7.m}{The Computing Profession}{Miscellaneous}{Ethics}
}

\teaser{
 \centering
 \vspace{2em}
 \includegraphics[width=\linewidth]{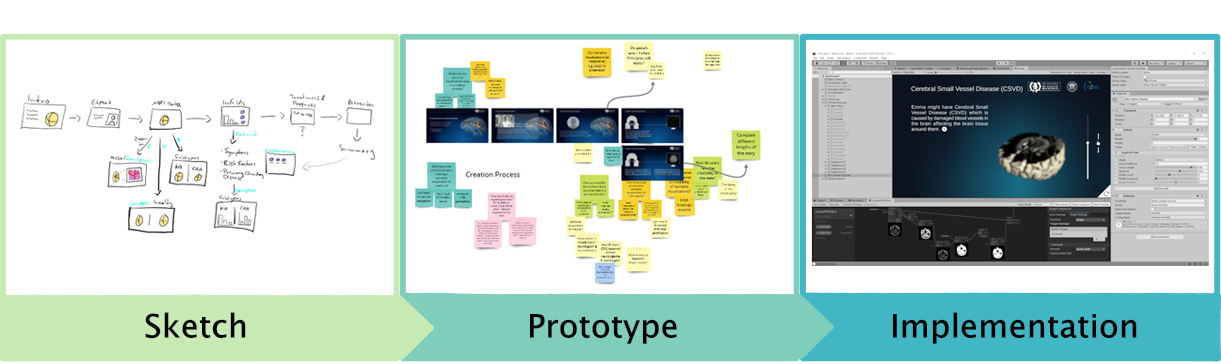}
 \caption{Main steps in developing our data-driven medical story about cerebral small vessel disease. We started by briefly sketching the story, followed by more concrete prototypes using Miro and finally the implementation in Unity.}
 \label{fig:teaser}
  \vspace{2em}
}

\vgtcinsertpkg

\begin{document}


\maketitle

\section{Introduction}

In the general public, there is a strong interest in medical topics such as the diagnosis and prevention of diseases as well as information on innovative treatment methods. These topics are often presented in a narrative format in popular TV broadcasts. Extended reports on recent research related to COVID-19 and a variety of YouTube videos for a wide range of health related topics also aim at general audiences~\cite{bora2018}. Moreover, government agencies, such as the World Health Organization inform the general public in a comprehensible manner on health-related developments.
Most of these narrative formats follow a linear structure where the only possible interaction is to move to the next or previous view. News outlets increasingly provide interactive data-driven stories with more advanced interaction possibilities in their online articles, e.g., to convey the spatio-temporal development of COVID-19 in terms of incidence, hospitalization and death as well as differences in terms of age, and gender.

Narrative visualizations became a mainstream research topic in the last decade~\cite{Ma2011,Meuschke2022}.
Thus, we build on a corpus of previous knowledge and experience and extend it to the design of narrative visualizations for communicating diseases, in particular neurological diseases. The communication of diseases provides unique possibilities for \textit{personalization}---an essential strategy to engage viewers~\cite{Bach2018}. Audience engagement can be increased by presenting an experienced physician diagnosing, treating, or researching a disease as well as by introducing an affected patient. Many diseases are characterized by specific locations with pathological changes. Thus, 3D visualizations that convey this spatial information are a potentially essential component of narrative visualizations in this area. For neurological diseases, such as epilepsy, brain tumors, cerebral aneurysms, and multiple sclerosis, this translates to displaying the brain with its different substructures, e.g., brain lobes as anatomic context for the location of the pathologies.

Narrative visualizations are designed for a variety of settings, including mobile devices, web-based systems and special settings that may involve large interactive displays. The latter may provide an immersive experience but are bound to a rather small set of users that are physically in a room which is equipped with the necessary and expensive device. 

Our work is motivated by the question of how neurological diseases can be communicated using narrative visualization techniques.
We focus on a web-based system to present our narrative visualization about cerebral small vessel disease to a general audience. In cooperation with our clinical and design partners we created a story that communicates \textit{preventable risk factors} as well as possible \textit{consequences} of this wide-spread but mostly unknown disease. This involved gathering data from clinical studies and visualizing these data in a way that enables interpretation by non-experts.
In summary, this paper makes the following contributions:
\begin{itemize}
    \item We propose a character-driven story structure for disease stories based on Campbell's Hero's Journey, see~\autoref{fig:campbell}. We describe in chronological order the events that a patient experiences when diagnosed with a disease.
    \item We describe our approach of adapting narrative components to address and communicate special characteristics of our medical data sets.
    \item We present and discuss lessons learned from generating a narrative visualization on cerebral small vessel disease which were developed from  a user study with 40 participants.
\end{itemize}


\section{Related Work}
The study of narrative medical visualization can be approached from two directions.
The first direction is to review work in the field of general narrative visualization and data story telling. We pay particular attention to developments in the direction of scientific visualizations and medical visualizations. The second direction is to analyze the science communication that is already being done by hospitals, health organizations and data journalists to educate people about medical topics.

\subsection{Background on Narrative Visualization}
\label{sec:background}


 Considerable work has been done on narrative genres \cite{Segel2010}, narrative structures \cite{hullman2011}, and the role of rhetoric to engage viewers \cite{hullman2013}. 

There are different concepts for designing a narrative like Campbell’s Hero’s Journey~\cite{campbell2008hero}, emotional arcs~\cite{Reagan2016}, the CFO (Claim, Facts, cOnclusion) pattern~\cite{kosara2017}, and Freytag’s Pyramid, which is based on Aristotle’s Three Act Structure~\cite{madej2008}. Brent Dykes' storytelling arc further derives from this Three Act Structure~\cite{dykes2019effective}. Yang et. al. discuss, how Freytag’s Pyramid can be applied to create structured data stories~\cite{Yang2022}.
Furthermore, a story can be told synchronously or asynchronously~\cite{lee2015more}. Synchronous stories are presented by a live narrator while asynchronous stories can be viewed by the user on their own.
Narrative visualizations cover a spectrum between being author-driven or reader-driven~\cite{Segel2010}. Author-driven stories combine a linear ordering of scenes with no interactivity to convey a clear message. With reader-driven stories, the author has only limited control of the user experience and the message they want to communicate. Where a narrative visualization is located on this spectrum is determined by the story structure. We focus on structures that follow a predefined main story branch, but offer interaction possibilities within, like the interactive slideshow~\cite{Segel2010}, as well as including optional side branches that can be visited additionally to the main story branch, like in the elastic structure~\cite{Seyser2018}. This conflict between user interaction and a pre-defined story is described as a \textit{narrative paradox}~\cite{Aylett1999}. 



Bach et al. suggest eighteen narrative design patterns that help to create a narration~\cite{Bach2018}. These include structuring the argumentation, creating a flow, evoke emotions, and strengthen user engagement. This body of knowledge was derived from various applications for information visualization, in particular for those that are based on tabular data, time-oriented data and to a smaller extent map-based data~\cite{Hullman2013_2, madhavan2012}. Such data exist, for example, in business and finance and thus a considerable commercial interest exists. 
Pictographs, introduced by the Austrian Otto Neurath in the 1930s to convey statistical data were used frequently since decades in a variety of contexts. These infographic elements are useful to represent, e.g., fractions. They are also invaluable  in communicating health-related data. Haroz et al. investigate the effect of using pictographs to convey information and could demonstrate that they are beneficial in terms of memorability and engagement~\cite{haroz2015isotype}. Being an important element in narrative visualizations infographics may be semi-automatically generated with \textsc{InfoNice} \cite{wang2018infonice}. 

Authoring tools to create visual data stories are another important area of research. Amini et al. analyze several data videos to suggest guidelines for an authoring tool suited for general audience~\cite{Amini2015}. In contrast, many authoring tools focus on static two dimensional visualizations following a linear story like ChartAccent~\cite{Ren2017}, Ellipsis~\cite{Satyanarayan2014}, and Data Comics~\cite{Zhao2015}.

There are large research gaps in the area of narrative visualization concerning  domains like scientific visualization\cite{Tong2018}. While there is some work done by NASA and in Science Museums~\cite{KwanLiuMa2012}, as well as in biological visualization~\cite{Kouril2021}, only little work has been done in the field of medical narrative visualization research.
Wohlfahrt and Hauser develop an approach for storytelling with volume visualizations, focusing on how to balance guidance and interactivity~\cite{wohlfart2007}. Although they employ medical volume data, they do not focus on informing about specific diseases and do not include other types of medical data such as 3D surfaces, slice images, or tabular data. So et al. mine social media data to create a story about biological, psychological, and social aspects of medical conditions for a general audience~\cite{So2021}, while the stories of Wohlfart and Hauser are suited to an audience that has at least basic domain knowledge.
A template for generating narrative medical visualizations according to disease data is suggested by Meuschke et al.~\cite{meuschke2021}. Their work is focused on how to present the main aspects of a disease comprehensively within a story. We build on this work in our design study about cerebral small vessel disease. However, our main goal is not a comprehensive explanation of the disease but to explore how telling a visual data story does aid conveying disease-specific messages that resonate and provide an effective call-to-action for a general audience?

\subsection{Science Outreach in Healthcare}
\label{sec:science_outreach}
Communication of complex medical findings to patients without a medical education is part of the daily routine of practioners. They inform about, e.g., illnesses, current medications, and vaccinations. More broadly, the field of public health is focused on the health of the general population. The World Health Organization (WHO) adapted the definition of public health from Acheson (1988): “the art and science of preventing disease, prolonging life and promoting health through the organized efforts of society” \cite{WHO}. 
Narrative visualizations can promote public health by enabling people to explore and better understand their own health independently. However, a distinction must be made between patient education and education of the general population.
When informing about illnesses, treatment options and prognosis are of particular interest for patients, while for the general population the focus is on prevention.

While many scientific areas, e.g., mathematics, astronomy, or philosophy feel abstract to a general audience, healthcare is a topic that affects people personally and emotionally. Thus, it is  especially important to communicate risks appropriately. We aim at presenting the risks in a way that motivates personal changes without scaring the users watching the story. 
However, several studies show that most people tend to underestimate personal risks~\cite{Welschen2012, Wachinger2012, Schmlzle2017}.

Communication in healthcare is important to enable self-determined decisions, e.g., by deciding for a specific treatment or preventing risk factors in their everyday life. Thus, many medical institutions are searching for ways to better communicate with their patients or with general audiences. The Charité in Berlin is working on using comics to improve patient education~\cite{charite}. At the University of Tübingen, students from medicine, biology, and rhetoric are working together to create visualizations for the scheduling and decision-making processes for medical procedures~\cite{tuebingen}. Apart from static 2D visualizations, there are also video formats for communicating medical topics. The german newspaper \textit{Focus} uses such videos to educate viewers about symptoms and prevention of common diseases, e.g., stroke~\cite{focus-stroke1, focus-stroke2}, and employ slices and 3D renderings of medical volume data.



\begin{figure}[tb]
 \centering
 \includegraphics[width=0.842\columnwidth]{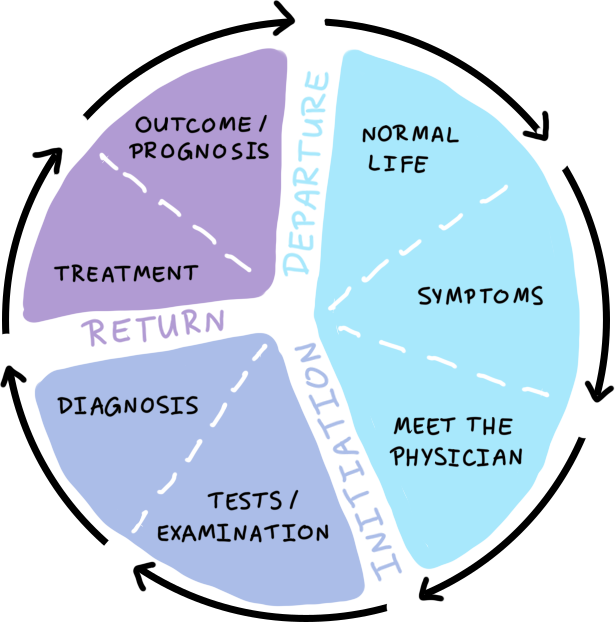}
 \caption{The patient's disease journey adapted from Campbell's Hero's Journey. During \textit{departure}, the patient has to leave their normal life because of symptoms and visits the physician. In the \textit{initiation} phase, the patient has to undergo examinations and tests to receive a diagnosis. Finally, during \textit{return} the patient receives treatment and a prognosis. The goal at the end is to return to normal life as much as possible.}
 \label{fig:campbell}
\end{figure}

\begin{figure*}[tb]
 \centering 
 \includegraphics[width=\textwidth]{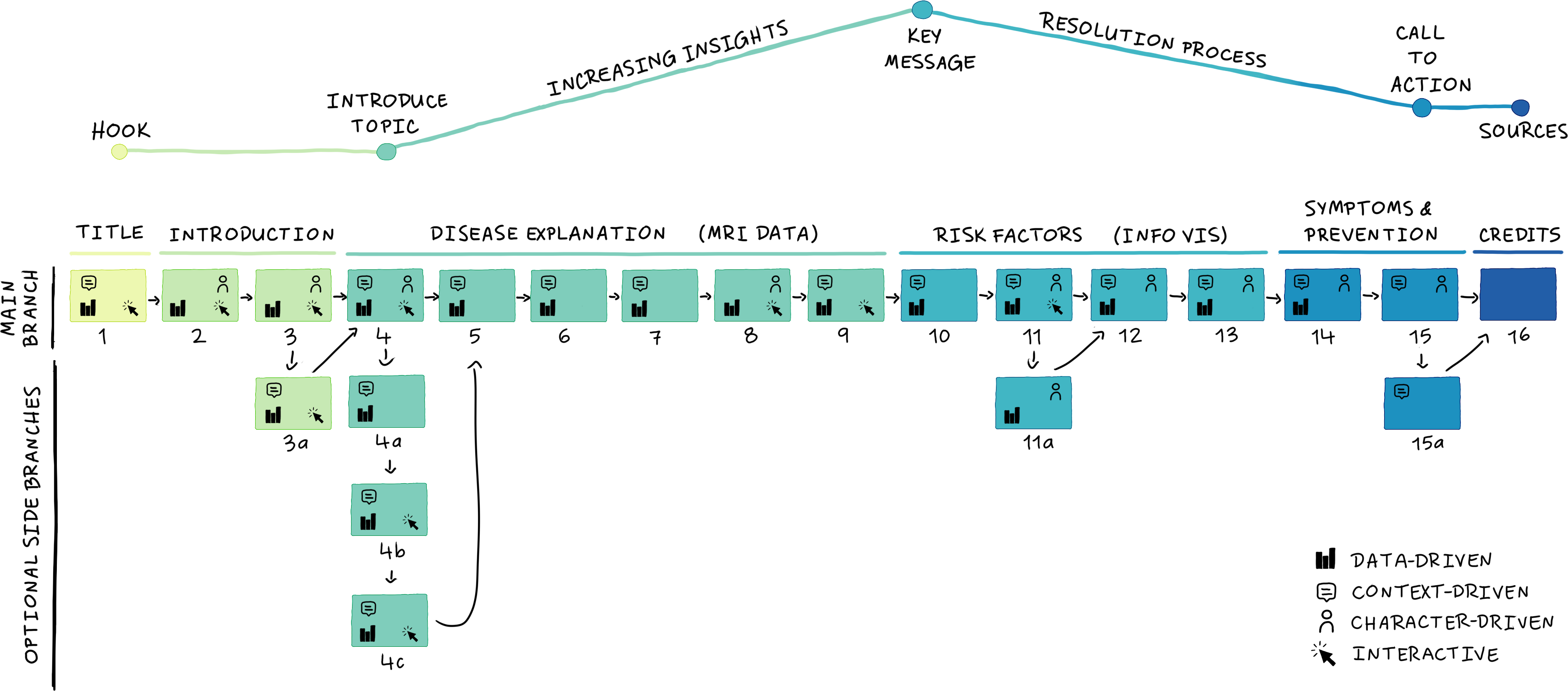}
 \caption{Our functional story structure mapped to an adapted version of Freytag's Pyramid and the storytelling arc. The story follows a linear main branch, offering to switch to optional side branches temporarily (so-called \emph{elastic structure}). The parts of our logical story structure are highlighted in different colors. In each slide we highlight whether there is data-driven, context-driven, character-driven, or interactive content.}
 \label{fig:arc_structure}
\end{figure*}

\section{Adapting Narrative Techniques for Storytelling of Diseases}
\label{sec:communicating}

\textit{Characters}, \textit{conflict}, \textit{content} and \textit{structure} are the main components of a visual data story~\cite{dykes2019effective, meuschke2021}. \textit{Characters} are not necessarily people, but can be objects. The \textit{characters} are confronted with a \textit{conflict} they have to solve. The \textit{content} of the story is built by introducing facts around this \textit{conflict}. The story is organized according to a narrative \textit{structure}, e.g., Freytag's Pyramid, Campbell's Hero's Journey or the CFO pattern.
The following research questions have emerged from exploring the design of a narrative story in the context of our  prototype:
\begin{itemize}
    \item To what extent must complex medical relationships be simplified in order to be understandable to a general audience without falsifying the underlying data and findings?
    \item What elements can help the audience better relate to the story?
    \item How can we usefully integrate 3D data into the story?
    \item Which character roles should be represented in the story and how should they be integrated?
    \item What level of interaction with medical data is appropriate for general audiences?
\end{itemize}

\subsection{Integrating Different Types of Content}
As the name implies, the main \textit{content} of a visual data story is the data itself. However, additional contextual information significantly enriches the story. 
Narrative visualization strives to engage the user with the story. To strengthen this engagement one or more \textit{characters} can be introduced whom the user can follow through the story. In the interest of data protection we use a fictional but representative patient as the main character.  Additionally, we introduce our clinical partner, who is a neurologist with advanced expertise in CSVD. 

When creating a story about a disease, the \textit{conflict} the character must face is being ill. In many cases, the journey of a patient fighting a disease has parallels to the three stages of the narrative \textit{structure} of Campbell's Hero's Journey: \includegraphics[height=0.7em]{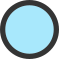}~\textbf{Departure}, \includegraphics[height=0.7em]{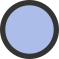}~\textbf{Initiation}, and \includegraphics[height=0.7em]{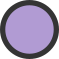}~\textbf{Return}, see~\autoref{fig:campbell}. During \includegraphics[height=0.7em]{figures/circle2_recolor.png}~\textbf{Departure}, the patient is forced to leave their \textit{normal life} due to occurring \textit{symptoms} of a (still unknown) disease. The patient then \textit{meets with a physician}, leading to the \includegraphics[height=0.7em]{figures/circle3_recolor.png}~\textbf{Initiation} stage. Next the patient must undergo different \textit{examinations and tests} to help physicians determine a \textit{diagnosis}. Afterwards, in the \includegraphics[height=0.7em]{figures/circle4_recolor.png}~\textbf{Return} stage, different \textit{treatment} options as well as a \textit{prognosis} are discussed. The overall goal is to return back to the patient's normal life as much as possible. In contrast to other narrative structures, as well as the template for disease stories~\cite{meuschke2021}, Campbell's Hero's Journey builds the story around the main character.

Based on these observations, we distinguish between three types of story content:
\begin{itemize}
    \item \textbf{Data-driven:} Content that is directly derived from the available data.
    \item \textbf{Context-driven:} Content that cannot be derived directly from the data but represents common (domain) knowledge or information aggregated from multiple studies, often provided by the domain expert.
    \item \textbf{Character-driven:} Content that is based on a real or fictional character, e.g., a domain expert or a person that is affected by the subject of the story like a patient.
\end{itemize}


\subsection{High Blood Pressure and CSVD: A Disease Story}
We now discuss the modification of these components to better serve our medical story about cerebral small vessel disease (CSVD).
Our main goal is to communicate to the audience that high blood pressure can lead to CSVD, which strongly increases the risk of dementia. This statement is based on contextual information and is supported by concrete data facts during the story. In~\autoref{fig:arc_structure} we show which slides of our narrative visualization are data-driven or context-driven.

CSVD describes a variety of conditions associated with damaged small blood vessels in the brain~\cite{Wardlaw2013}. These pathological blood vessels cause damage in the brain tissue. In the long run, CSVD can lead to motor and cognitive impairment, and is a high risk factor for dementia. CSVD occurs 6-10 times more frequently than stroke and is the most common incidental finding on brain scans~\cite{Chojdakukasiewicz2021}. There are two subtpes of CSVD, called hypertensive arteriopathy (HA) and cerebral amyloid angiopathy (CAA). Depending on the subtype, the treatment of CSVD differs greatly. While affecting a large part of the population, it is unknown to many people. This makes it an interesting topic for scientific outreach. Furthermore, the prevention of CSVD can be done easily on an individual level by aiming for a healthy lifestyle and regularly checking the personal blood pressure.
However, the disease is not very well-understood and there are still many unanswered questions. We worked closely together with our clinical partner to present up-to-date information that is in line with the scientific consensus.

While we have shown in~\autoref{sec:science_outreach} that there are many artistic media, e.g., cinematic videos and hand-crafted comics, used to communicate medical topics, our aim is to mainly present real data. Using real data also enables non-artists, such as domain experts doing science communication, to create narrative visualizations. 
However, using real data presents us with the challenge of combining medical volume data with tabular data and domain knowledge into an appropriate visualization. 
Furthermore, the study data to which we have access does not cover all aspects necessary to be communicated. In some cases, the study data may not be representative for the scientific consensus concerning CSVD. 
Our clinical partner helped to review the data and discuss optimal presentation approaches.
Compared to most other narrative visualization examples our story is able to show more details which we derived from real medical data.
Due to the amount of content, the story is divided in several logical parts which represent semi-independent sub-stories inside the overall story leading to a \textit{nested narrative visualization structure}, see~\autoref{fig:arc_structure}. 
The same can be noticed considering the genres from Segel and Heer which range from an annotated chart to an interactive slideshow~\cite{Segel2010}. However, it often happens that a slideshow contains (annotated) charts. Therefore, stacking sub-stories different genres is not uncommon. 

\begin{figure*}[tb]
 \centering 
 \includegraphics[width=\textwidth]{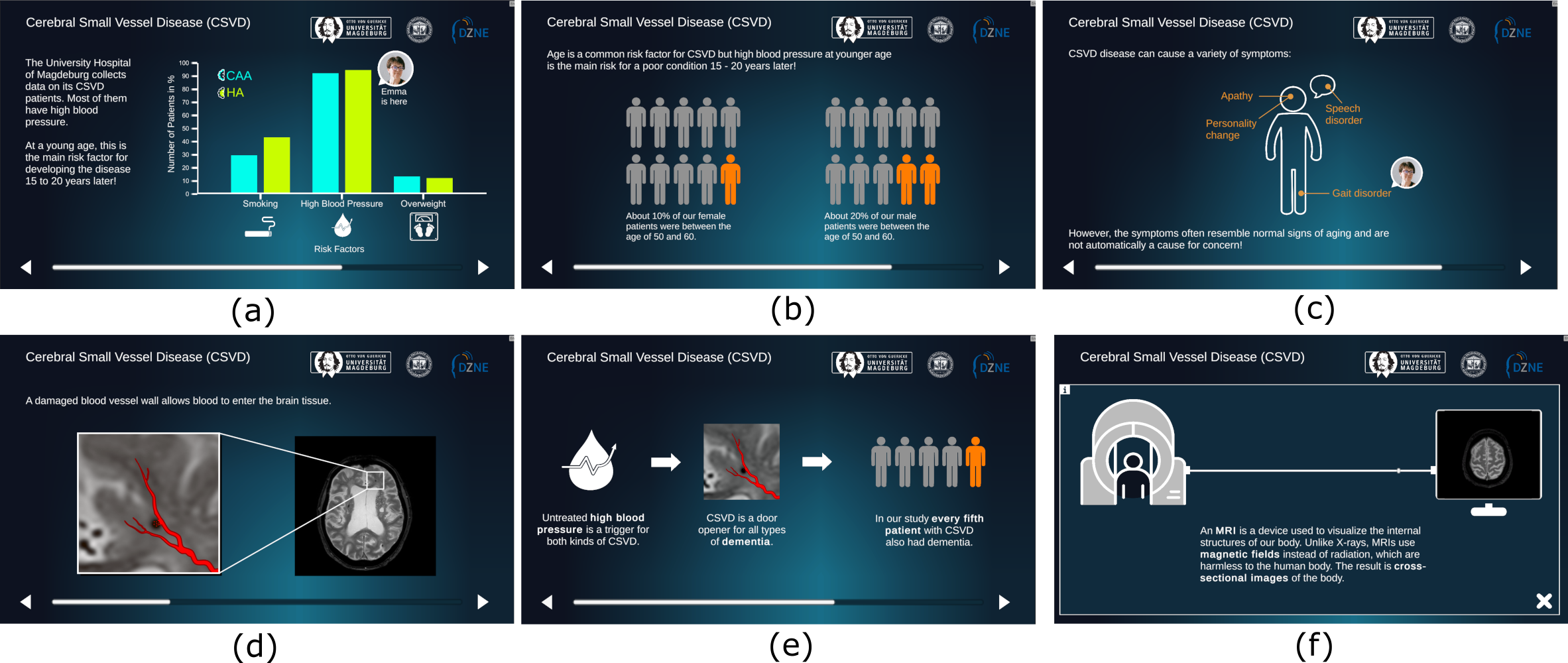}
 \caption{Selected slides from our visual data story, showing (a) distribution of risk factors, (b) the distribution of patients regarding age, (c) pictographs of the frequency of symptoms, (d) schematic vessel drawings, (e) the visualization of our key message, and (f) the interactive explanation of the MRI. }
 \label{fig:slides}
\end{figure*}

\section{Story Design}
Designing the story was a highly interdisciplinary work between computer scientists, medical experts an interaction designer and a medical illustrator.
We identified several workflow steps, that were important to develop our medical narrative visualization. 

\paragraph{Preparation.}
Story design first requires a number of preparation steps.  At first, we found a domain expert who then decided on a topic which is interesting for a general audience and defined a target audience. Next, data must be acquired. The domain expert may be able to facilitate this, else, depending on the topic, open source data repositories may be available, such as the NIH Cancer Imaging Archive\footnote{https://www.cancerimagingarchive.net/}.
In our case, we decided for the topic of the wide-spread CSVD. We cooperated with a neurologist who is actively researching the disease and provided us with MRI data as well as tabular data from cross sectional patient studies. The data was processed by segmenting the damaged brain areas in the MR images, generating surface models of volume data and filtering the tabular data according to the information we want to present during the story. Our target audience is adults from the age of 45, because this is the critical age to focus on prevention of CSVD.

\paragraph{Narrative Structure.}
We then transformed the data into a story by arranging it in a narrative structure and adding context-driven information. To decide which data is needed, the narrative intents regarding the target audience have to be identified. Narrative intents include informing about the disease and the key insights a user should recognize in the course of the story as well as the resulting call-to-action. 
This results in our key message that \textit{high blood pressure can lead to cerebral small vessel disease, which strongly increases the risk of dementia}. The corresponding call-to-action is a change to a healthier lifestyle and regular monitoring and, if necessary, treatment of high blood pressure.


When designing our narrative, we referred to Freytag's Pyramid and Brent Dyke's Storytelling Arc, see~\autoref{fig:arc_structure}. In contrast to the storytelling arc, we placed the \includegraphics[height=0.7em]{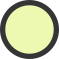} \textbf{hook} at the very beginning. Since our story is told asynchronously, its title screen needs to be motivating for people to start watching it. We continue by introducing our characters, which are the medical expert as well as the patient.
We want to communicate to users with similar risk factors that in a few years, they may find themselves in the role of this patient. Thus, a large part of our visual data story is character-driven. Freytag's Pyramid describes the structure of a story in terms of its tension. Thus, we additionally used an adapted version of Campbell’s Hero’s Journey to structure the character-driven content describing the journey of our patient, see Fig.~\ref{fig:campbell}. We focussed on its parts about \textit{diagnosis}, \textit{treatment} (which in our case includes secondary prevention) and \textit{prognosis} to  emphasize the key message.

To \includegraphics[height=0.7em]{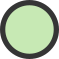} \textbf{introduce the topic}, we describe the disease with the aid of real MRI data. Using real data provides a more complete picture and \includegraphics[height=0.7em]{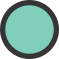} \textbf{increases the insight} of the users into its background. 
We present data from a cross-sectional study using easy-to-understand information visualization like bar charts, showing the distribution of different risk factors leading to our \includegraphics[height=0.7em]{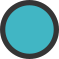} \textbf{key message}. 
Showing symptoms and risk factors is part of the \includegraphics[height=0.7em]{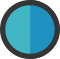} \textbf{resolution process}, which leads to the  \includegraphics[height=0.7em]{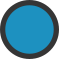} \textbf{call to action} at the end of the story. The story guides users through the various strategies that they can use to mitigate their risk for the disease. These strategies are summarized at the end of the story. During the resolution process, we included our fictional patient frequently by showing which of the presented risk factors and symptoms apply to her. We do this in order to create empathy between the user and the data presented. 
Similar to the disease template of Meuschke et al.~\cite{meuschke2021}, we presented methods for the prevention of disease at the end of our story, to emphasize the call-to-action. However, we did not follow the strictly sequential structure of the template but instead foreshadowed and repeated some information, such as symptoms, risk factors and prevention. This foreshadowing and repetition of elements is designed to capture user attention and to cement key insights into memory

The last slide contains further information about the creators of the story, similar to a credits screens in movies, showing the \includegraphics[height=0.7em]{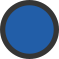} \textbf{sources} of information to strengthen the credibility of the data.

\begin{figure}[tb]
 \centering 
 \includegraphics[width=\columnwidth]{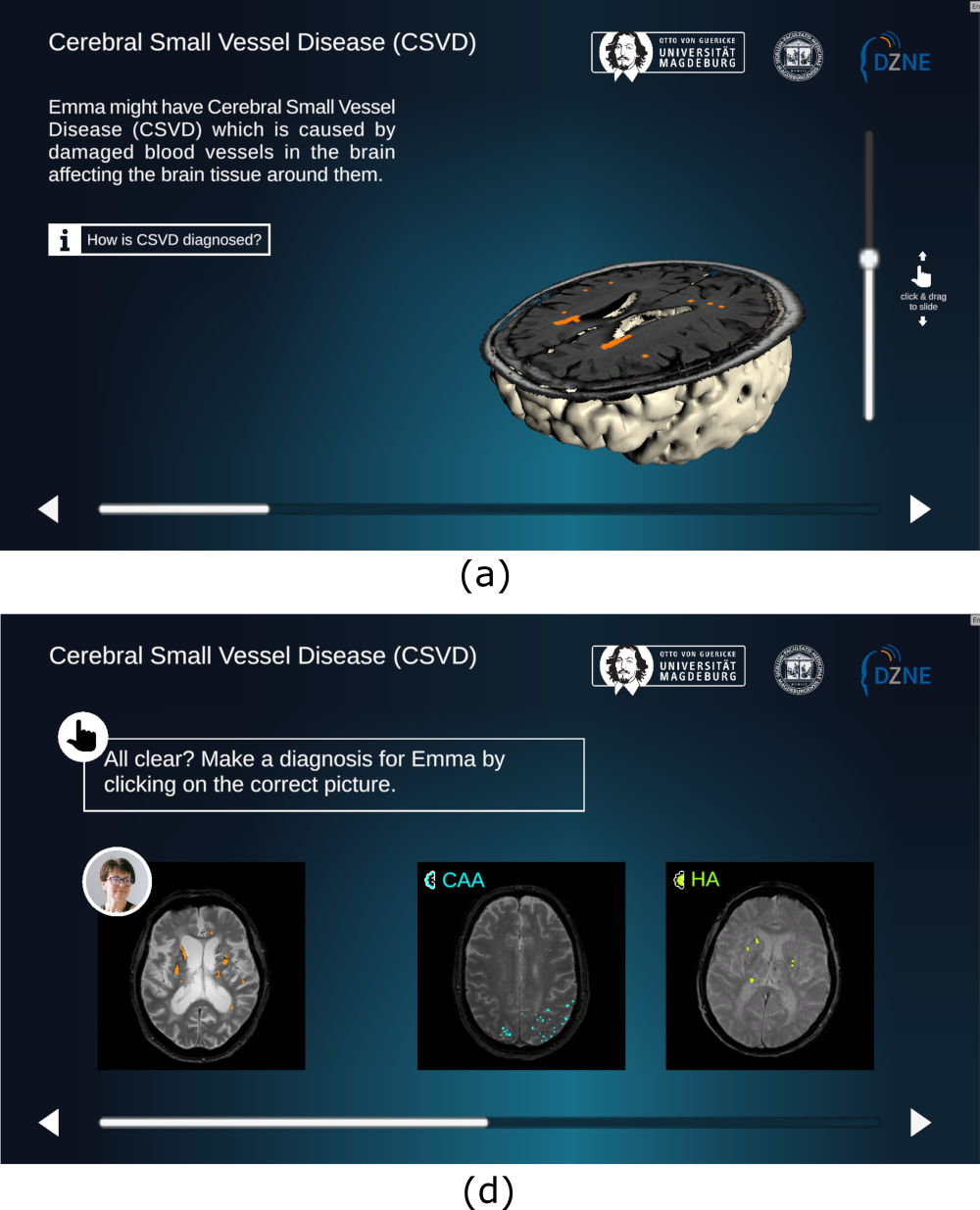}
 \caption{Two screenshots from our narrative visualization, showing slides containing (a) 3D interaction possibilities (rotating and clipping) and (b) a point and click interaction task.}
 \label{fig:interaction}
\end{figure}

\paragraph{Design.}
Our interdisciplinary team consisted of visualization researchers, a medical domain expert on CSVD, as well as a certified medical illustrator. We used the online whiteboard Miro\footnote{Miro: \textit{www.miro.com}, Amsterdam NL and San Francisco U.S.} to create a story board and discuss different layout proposals, see~\autoref{fig:teaser}. Miro offers a great flexibility: Images, text, as well as graphical elements like arrows can be arranged freely, making it suitable for cooperative work between all stakeholders.

We scheduled regular design meetings to enable an iterative feedback loop. In these sessions, we discussed, e.g., the general visual design, use of color and interface elements. We decided on a dark background for optimal contrast in vessel visualization, and subsequently added white text and intense accent colors. Each sub-type of CSVD was assigned one color, and this classification was applied consistently throughout charts, pictopraphs and visualizations of damaged brain tissue. Furthermore, one additional accent color was defined to highlight damaged brain tissue not assigned to the two subtypes. Each color was chosen to provide sufficient contrast in all areas of application. For the interface elements, we decided on a clean flat design that would not draw attention away from the main story elements, i.e. the brain visualizations.

We decided to use a bar chart because this was the best encoding for the data and is frequently employed in narrative visualization due to their simplicity. To support its interpretation we added short textual descriptions, see~\autoref{fig:slides} (a).
Additionally, we employ \textit{pictographs} (recall \cite{haroz2015isotype}) suitable to convey health data. We chose to use the silhouettes of a stylized human, proportionately colored to represent the proportion of patients from our clinical study data experiencing risk factors, see~\autoref{fig:slides} (b) and (e). 
To explain the origin of the disease, we used cross sectional images of the brain and included sketches of the damaged blood vessels in realistic positions next to the damaged brain tissue, see~\autoref{fig:slides} (d).

To visualize facts for which we do not have tabular data nor volume data, we have decided to insert icons, see~\autoref{fig:slides} (e). These were also used to show risk factors on the x-axis on relevant bar charts, see~\autoref{fig:slides} (a). To create the icons, we used the open-source vector graphics editor Inkscape\footnote{Inkscape: https://inkscape.org}. We kept the icons as simple as possible and made them monochrome so they would not interfere with our accent colors. We made one exception, since the sub-types of the disease have long complex names in medical terminology and are not easily paraphrased, we paired their acronyms with icons. These icons are partly colored in the respective accent colors blue and green, e.g., see~\autoref{fig:slides} (a).

At the beginning of the story, we introduce our medical expert and our patient using photos to give users a better impression of the characters of our story. We used a stock photo of a 60-year-old woman. This fits the profile of someone with high risk for CSVD and is the typical age of onset for this disease.
We also included an icon representing the patient on many slides to relate the data to her, see~\autoref{fig:slides} (a) and (c). This icon consists of a small cropped photo of our patient and marks which data is patient-related, e.g., which of the presented risk factors and symptoms apply to her.

We chose the narrative genre of an interactive slideshow. To make transitions between the different slides smoother, we created a \textit{fade in} and \textit{fade out} effect. To navigate between the slides, we included a forward and a backward button on the bottom of our slides. Between them, a progress bar shows how much of the story the user has already viewed. It serves to motivate the user to continue the story and show how far they have progressed. An icon is displayed in the description text that drives the narrative, if further information is available to the current topic, see~\autoref{fig:interaction} (a). Clicking on it opens optional slides.
In the header we display the topic of our story, as well as the logos of the contributing organizations, to lend credibility to the information presented.

Some slides have \textit{interaction possibilities} which trigger \textit{animations} within the slide. We introduce a different set of \textit{icons} to highlight different interaction possibilities, see \autoref{fig:interaction}. 
We only allow rotation around the up-axis to simplify the user interaction. Some 3D models can be clipped to visualize the damaged brain tissue within. We chose a horizontal \textit{clipping plane} to give the user a simple way of viewing the different MRI slices along one axis without loosing the anatomical context. This interaction is inspired by the way medical medical experts navigate through volume data. Therefore, the user can steer the plane by using a slider. From the slider position we project the corresponding MR image on the clipping plane. To emphasize this interaction, we included a hand icon paired with one up and one down facing arrow to show that the user can adjust the slider knob upwards and downwards.
Some slides contain individual tasks where the user has to point and click or grab and move elements on the screen to solve tasks. In these cases, a hand icon combined with a short \textit{instruction} is shown. For example, we explain that the cross sectional slices that are common results of medical imaging, by letting the user grab a pictograph of a patient and moving it through a MRI. Based on the position of the pictograph, a corresponding slice image of the brain and explaining text is shown, see~\autoref{fig:slides} (f). Though this does not depict the imaging process in the abstract, it allows the user to interactively explore and understand how medical volume data is aquired.

\paragraph{Implementation.}
In contrast to our nested narrative visualization model, which presents the logical structure our story is following, the implementation follows a \textit{functional story structure}, see~\autoref{fig:arc_structure}.
The functional structure describes, how the user can navigate through the story, highlighting the main story branch, optional side branches and animation steps. We decided for an \emph{elastic structure} where the users have to follow the main story branch but also have the possibility to temporarily switch to small optional branches.

We chose the Unity\footnote{Unity Technologies: https://unity.com, San Francisco U.S.} game engine to implement our narrative visualization. Unity is a powerful tool which offers several advantages. Many pre-made functionalities to visualize 2D as well as 3D elements are available. The Unity shader graph allows for rapid creation of create different materials for both 2D and 3D. 
Furthermore, Unity comes with a visual editor, allowing to interactively place and move objects on the screen, create individual animations, implement user input as well as interaction possibilities, and change many other settings, e.g., supported languages. This makes the design process much easier.
In addition, Unity allows for export of the final application to almost every platform, including different operating systems, mobile as well as web applications, and even gaming consoles. This way, the same application can be deployed for different output media without large implementation effort.
We chose a WebGL export and uploaded the application on our website to be able to reach as many people as possible. It can be viewed at \texttt{https://visualstories.cs.ovgu.de/stories/csvd-en/}.

\section{Evaluation and Results}
To evaluate our narrative visualization, we used the free online survey tool LimeSurvey. The survey was anonymous, with questions regarding personal risk factors being voluntary. In addition, we had the questionnaire approved by the university's data protection officer. We gave the survey to several people at the same time to mitigate the risk of backtracking and the survey ran for one week. 
Participants were recruited from our personal environment, taking care to represent a wide age range.
The participants were asked to do the survey directly after viewing the story.
Survey and story are provided as supplemental material. Our goal is to get feedback on how stories should be designed in the future and how the current story should be improved to answer our research questions in \autoref{sec:communicating}. Therefore, we evaluated our decisions concerning \textit{design}, \textit{navigation} and \textit{interaction} techniques as well as \textit{information} included in the story with regard to their influence on the \textit{credibility}, \textit{risk perception}, \textit{memorability}~\cite{Borkin2016}, \textit{engagement}~\cite{Boy2015, Amini2018}, as well as \textit{appeal} and \textit{comprehensibility}~\cite{Figueiras2014} of our story.
We focused on the following aspects:
\begin{itemize}
    \item \textbf{Credibility of the Data.} If the data presented in the story is not credible, users will not consider the content to be serious~\cite{dykes2019effective}. We asked if the participants think that the information shown during the story was credible and if they believe to know the origin of the data as well as the story author. They further had the possibility to name aspects of the story that influenced their trust.
    \item \textbf{Risk Perception and Call-to-action.} To evaluate whether participants are responding to our call-to-action to minimize risk factors for CSVD, we asked about their personal risk assessment and if they felt motivated to change their lifestyle after seeing the story. Questions regarding personal risk factors as well as risk assessment were not mandatory in the survey.
    \item \textbf{Memorability.} Narrative visualizations strive to enhance the memorability of the presented information~\cite{Borkin2016}. To evaluate, how well the participants memorize key aspects of our story, we designed a small quiz comprising single choice as well as multiple choice questions.
    \item \textbf{Story Structure and Engagement.} The engagement of a user is an important goal of storytelling and one of the five narrative pattern groups by Bach et al.~\cite{Bach2018}. We evaluated the engagement of the participants by asking them to rate different statements, e.g., if the story triggered their emotions, if they felt involved, if focusing on the story was easy and if the story was entertaining. To see if the length of the story influenced engagement, participants were further asked if the story was too long, if they skipped parts or if they viewed the optional additional information.
    \item \textbf{Visualizations and Design.} We included various different 2D and 3D visualizations, as well as icons and asked the participants to rate them according to their appeal and comprehensibility, to evaluate if they are suitable for an audience of non-experts. 
\end{itemize}

We recruited a group of 40 people representing a general audience.
The participants are between 18 and 83 years old with an average age of 42.36. 22 of them are female, 17 are male and 1 is non-binary. Since our target audience are people from the age of 45, which applied to 19 participants, we investigated whether there were differences in the evaluation of the visualizations between people in this group and the younger participants. However, we did not observe any striking differences in the results. 

\begin{figure}[tb]
 \centering
 \includegraphics[width=\columnwidth]{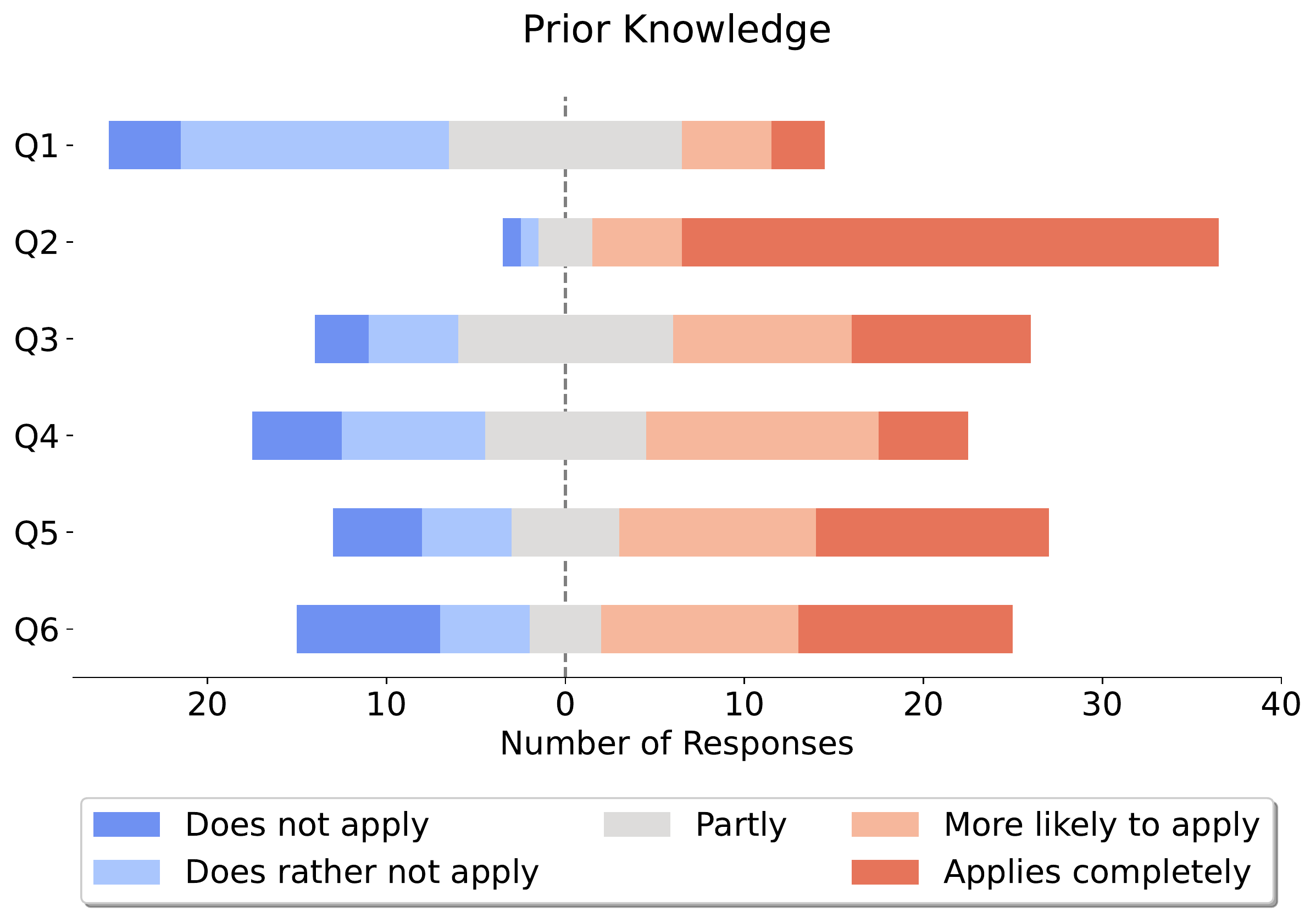}
 \caption{Evaluation results regarding the prior knowledge of our study participants. Participants were asked to rate the following statements: (Q1) I have good medical knowledge. (Q2) I am familiar with diagrams, e.g., bar chart or line chart. (Q3) I am familiar with medical illustrations, e.g., from textbooks. (Q4) I am familiar with medical images, e.g., from MRI, CT or X-ray. (Q5) I am familiar with 3D visualizations. (Q6) I am familiar with interactions with 3D visualizations, e.g., rotating.}
 \label{fig:preexpirience}
\end{figure}

To better interpret the answers, the participants were asked to self-assess their prior knowledge in medicine and visualization at the the beginning of the survey, see~\autoref{fig:preexpirience}. 20\% of the participants stated that they have good or very good medical knowledge and are also professionally working on medical related topics. 
87.5\% of participants reported familiarity with bar charts and line graphs. Half of the participants agreed or strongly agreed that they are familiar with medical illustrations (25\% and 25\%, respectively) and medical images (32.5\% and 12.5\%).
60\% of the participants stated that they are familiar with 3D visualizations (27.5\% agreed, 32.5\% strongly agreed) and 57.5\% are also familiar with 3D interactions (27.5\% agreed, 30\%~strongly~agreed). In summary, the prior knowledge of our participants have more prior knowledge in the field of visualization than prior medical knowledge.

\begin{figure}[tb]
 \centering
 \includegraphics[width=\columnwidth, clip, trim = {0 1.5cm 0 0}]{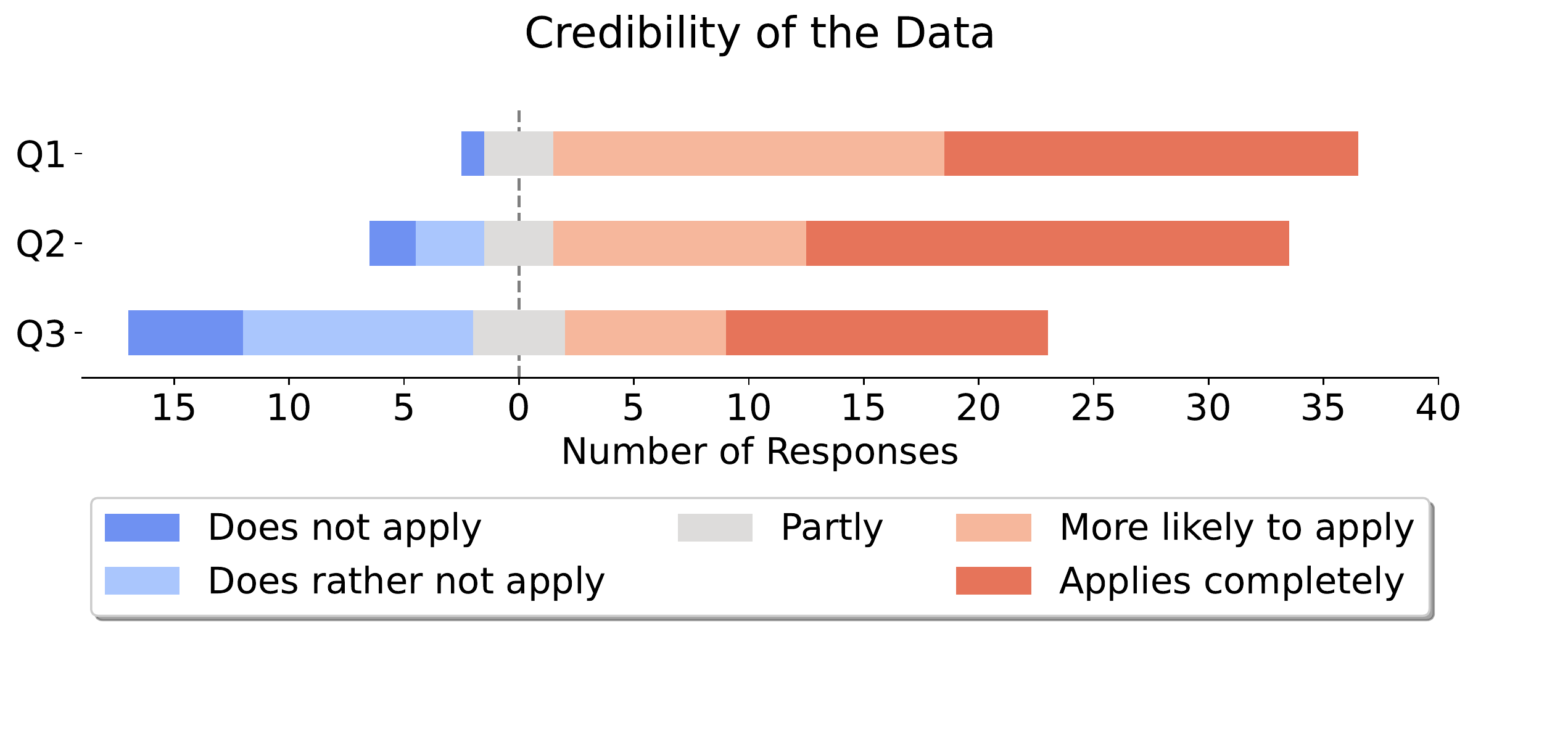}
 \caption{Evaluation results regarding the credibility of the data. The following questions were asked: (Q1) I consider the information shown to be trustworthy. (Q2) In the story, it was clearly shown from whom the information for the basis of the visualization came. (Q3) It is clear to me who is the author of the story.}
 \label{fig:credibility}
\end{figure}

\paragraph{Credibility of the Data.}
The credibility of the data was rated generally high by the users, see~\autoref{fig:credibility}. 45\% strongly agreed that they think the presented information is trustworthy and 52\% strongly agreed that the story showed the source of the data. Participants had the option to name aspects that improved or impaired the credibility of the story. Positive aspects were the \textbf{logos of trusted institutions} like the University and involved medical institutions, the \textbf{introduction of our domain expert} and the \textbf{professional looking design} of our story. Trust was also strengthened by the fact that the \textbf{source of the data is known} and the information is \textbf{presented in a clear and plausible manner}.
Negative remarks were that the authors are not named directly and there was \textbf{no contact opportunity}. Some participants mentioned \textbf{missing meta information} about the study data and that there are no concrete sources of the information, e.g., external links. Furthermore, some users experienced \textbf{technical problems} during the presentation such as images overlapping.

\paragraph{Risk Perception and Call-to-action.}
47.5 \% of the participants stated that they do not believe they have any risk factors for CSVD and 20\% were not sure. The two most common risk factors are high blood pressure (25\%) and being overweight (20\%). Only 5\% of participants believe that they have a high risk of getting CSVD while most believe they have a low (60\%) or very low risk (12.5\%) and 5\% did not answer the question. 20\% of the participants said that they want to change their lifestyle after seeing the story. 

\paragraph{Memorability.}
We asked the participants several content-related questions to see how well certain facts were remembered. Therefore, we asked a series of single choice as well as multiple choice questions. Our key message that high blood pressure is the most important risk factor (correctly answered by 90\%) and that CSVD leads to dementia (correctly answered by 95\%) were memorized by the majority. 

We further asked what happens if the brain is affected by CSVD. Most of the participants were also able to reproduce this information correctly. More participants remembered that clogged blood vessels prevent supply to some brain areas (correctly answered by 87\%) than that damaged vessel walls lead to blood entering the brain tissue (correctly answered by 77\%) and that both damages the brain tissue (correctly answered by 62.5\%).
Overall the prevention methods were remembered well by the participants. Most recalled that  regular exercise (97.5\%), aiming for normal weight (87.5\%), stop smoking (90\%), checking and regulating the blood pressure (82.5\%), as well as regular and sufficient sleep (95\%). 
Only 55\% stated that the reduction of alcohol consumption is a way to reduce the risk of getting CSVD. The false answers that drinking 2-3 liters a day and avoid sitting also reduce the risk, were only selected by 7.5\% each.

We introduced two subtypes of CSVD, which were only presented by their acronyms. The participants had the task to choose the two correct options of a series of similar looking acronyms to see how well these abstract names were remembered. HA was correctly chosen by 87.5\% and CAA by 65\% of the participants. 30\% or less chose incorrectly.

\begin{figure}[tb]
 \centering
 \includegraphics[width=\columnwidth]{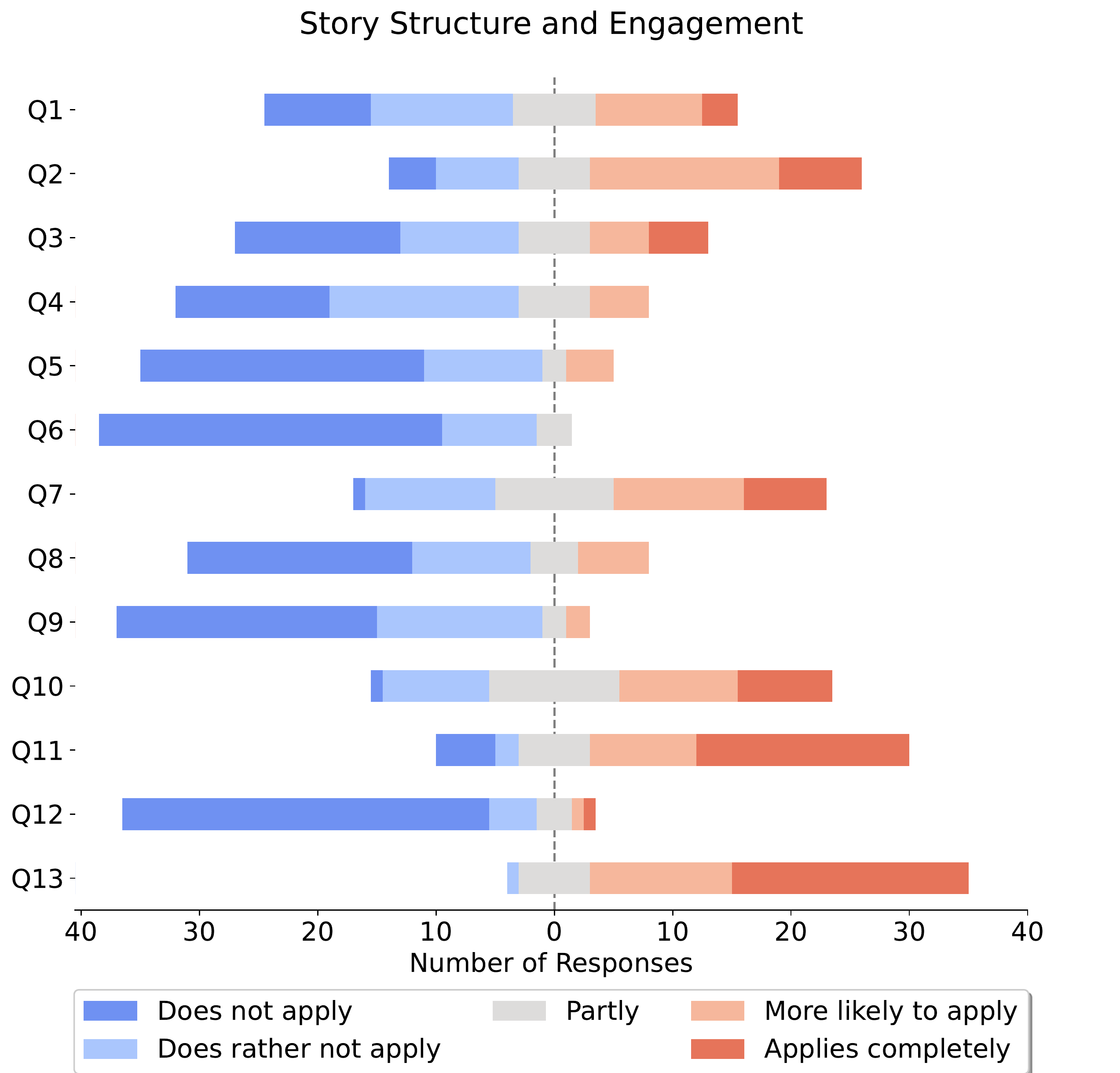}
 \caption{Evaluation results regarding the engagement and structure of the story. The following questions were asked: (Q1) The story triggered my emotions. (Q2) I felt actively involved in the story. (Q3) I lost track of time while watching the story. (Q4) While watching the story, my mind wandered. (Q5) I had a hard time focusing on the story. (Q6) I would have preferred to end the story early. (Q7) The story was entertaining.
(Q8) I found the story boring. (Q9) The story was too long. (Q10) I would recommend my friends to watch this story. (Q11) I looked at the additional information in the story. (Q12) I skipped parts of the story. (Q13) The additional content complemented the story well.}
 \label{fig:engagement}
\end{figure}

\paragraph{Story Structure and Engagement.}
We investigated the engagement of the users in the story, see~\autoref{fig:engagement}. We also asked how the structure of the story was perceived by the participants to assess, if there are aspects that disturbed the engagement. Furthermore, we measured the time it took the participants from receiving the link to the story until finishing it and starting the survey. Four of the times were unrealistically short (between 5 and 51 seconds) and also much shorter then the other values. We think that these times were caused by participants closing the survey tab while watching the story and reopening it afterwards, leading to the time not being measured correctly. Most of the other participants spent between 2:49 minutes and 13:54 minutes seconds viewing the story with an average of 8:87 minutes. Two participants took more time staying 19:59 minutes and 24:56 minutes in the story, possibly being interrupted by other tasks.

The majority of participants disagreed (35\%) or strongly disagreed (55\%) that the story was too long and 67.5\% agreed or strongly agreed that they watched the additional information. 80\% agreed or strongly agreed that the optional content complements the story well. In contrast, 5\% agreed or strongly agreed they had skipped parts of the story. Four of the participants gave the feedback that they were \textbf{lost between the different story branches} and were not sure if they are missing parts of the story. Sometimes it was \textbf{not clear that elements on the slides are clickable} to show additional information. In addition, the optional information also led to \textbf{gaps in the narrative flow} of the story.

We further asked the participants if they found the story entertaining (27.5\% agreed, 17.5\% strongly agreed) or if they were bored while watching it (15\% agreed, 0\% strongly agreed). 30\% of the participants indicated that the story triggered their emotions and many of would recommend the story to their friends (25\% agreed, 20\% strongly agreed). A majority of the participants stated that they felt engaged in the story (40\% agreed, 17.5\% strongly agreed).

\begin{figure}[tb]
 \centering
 \includegraphics[width=\columnwidth, clip, trim = {0 3.3cm 0 0}]{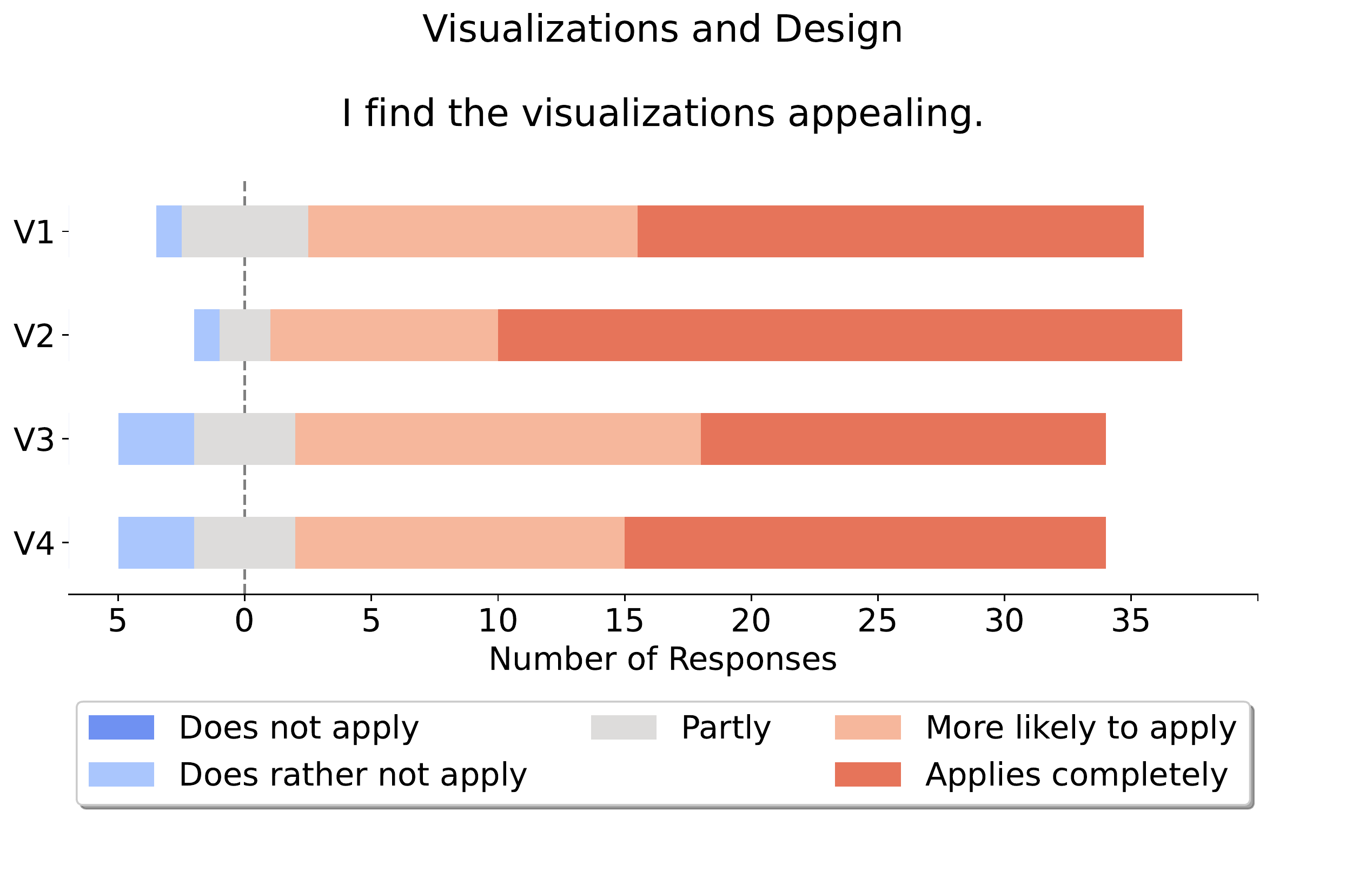}
 \vspace{0.2cm}
 
 \includegraphics[width=\columnwidth, clip, trim = {0 1.1cm 0 1.5cm}]{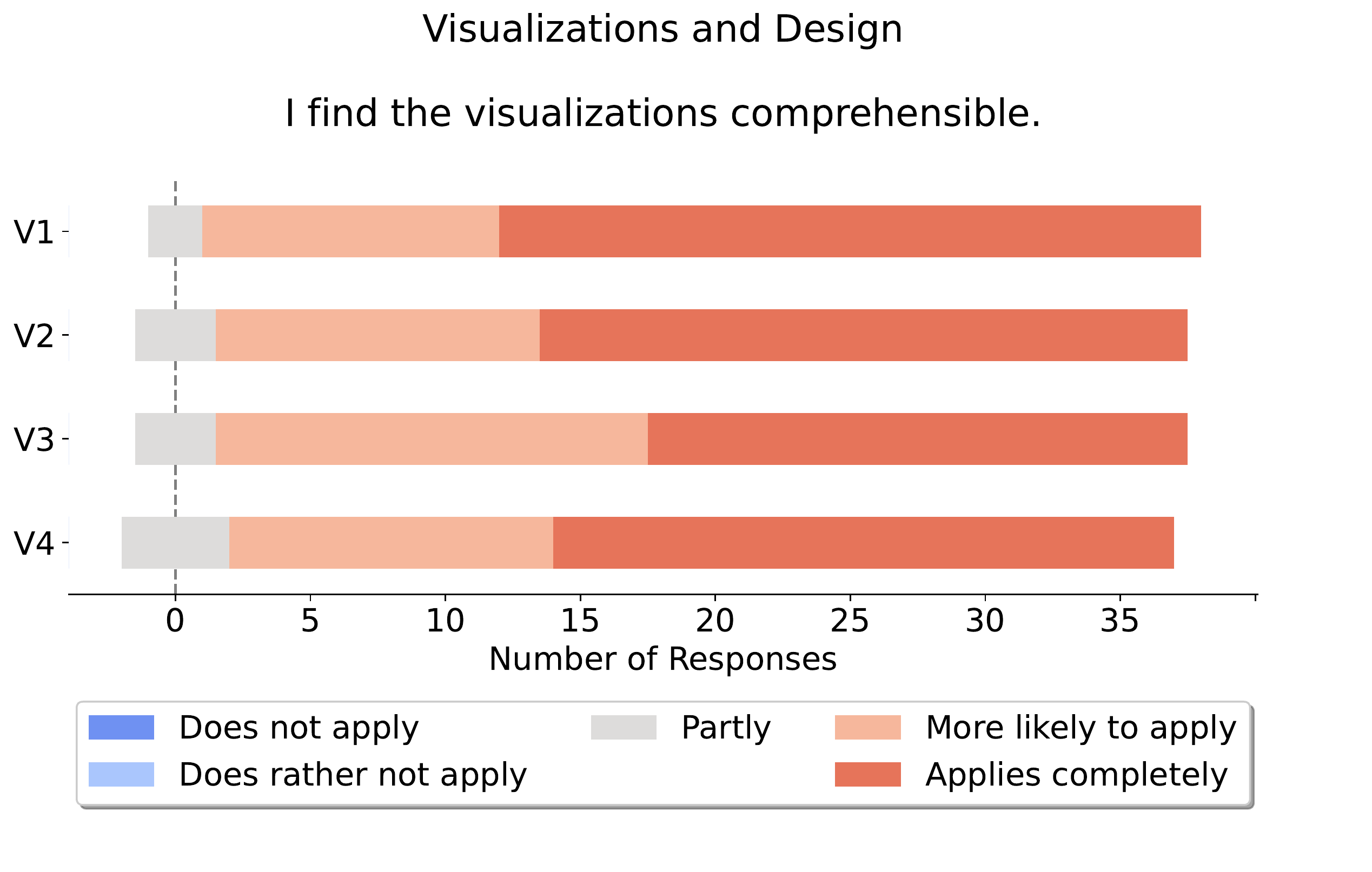}
 \caption{Evaluation results regarding the appeal and comprehensibility of our different visualizations: (V1) 2D MR images of the brain, (V2) 3D model of the brain, (V3) Symbols/Icons, (V4) Color choice.}
 \label{fig:appeal_comprehensability}
\end{figure}

\paragraph{Visualizations and Design.}
We asked the participants if they found the visualization in the story appealing and comprehensive, see~\autoref{fig:appeal_comprehensability}. The color choice was appealing for the majority of participants (32.5\% agreed, 47.5\% strongly agreed). The same applies to the medical visualizations of the 2D MR images (32.5\% agreed, 50\% strongly agreed) and 3D models of the brain (22.5\% agreed, 67.5\% strongly agreed). 40\% of the participants agreed and 40\% agreed strongly that the icons were appealing.

Furthermore, the comprehensibility was rated positively for the MR images (27.5\% agreed, 65\% strongly agreed ), 3D brain models (30\% agreed, 60\% strongly agreed), icons  (40\% agreed, 50\% strongly agreed) and choice of color (30\% agreed, 57.5\% strongly agreed), see~\autoref{fig:appeal_comprehensability}. There was no major difference in the results of participants who stated to have prior experience with medical visualizations compared to the ones with no prior experience.

\section{Discussion}
In the following we want to reflect on our work and discuss our evaluation results as well as the transferability of our approach to communicate neurological diseases in general.

\subsection{Discussion of Evaluation Results}
Our target group are people aged 45 or older. Therefore, we took special interest in how the evaluation results of this group differ from the results of our younger participants. However, we could not find large discrepancies between older and younger participants concerning the interaction and navigation in the story. We believe, that this is due to the similarity of our application to commonly known media products like slideshows or e-books. Our \textbf{interaction possibilities were designed regarding simplicity, thus the participants were not required to have any specialized preliminary knowledge}. Thus, younger participants that might have more experience in interaction with 3D models, e.g., from video games, did not have a major advantage.

Furthermore, we could not find any striking difference in the rating of the comprehensibility of the visualizations between participants with and without previous knowledge with medical visualizations. We conclude that this is due to the \textbf{simplified presentation of the data}, e.g., the highlighting of damaged brain tissue in MR images using colors and the language without jargon.

Concerning the risk assessment, 25\% of our participants believe that they have high blood pressure, however, only 5\% rate their risk of developing the disease as high. Even though we presented high blood pressure as the most important risk factor for CSVD, people might neglect its impact due to the amount of several other risk factors. Thus, having only one risk factor might not seem to raise the risk too much. Therefore, \textbf{a clearer distinction between risk factors of different importance needs to be presented}, e.g., by showing how much each risk factor contributes to the personal risk of CSVD. Another alternative would be to let people select the risk factors that apply to them during the story, and generate personally customized feedback. However, it is a typical phenomenon that people underestimate personal risk~\cite{Welschen2012}.

Our participants rated the information presented as credible due to the presentation of logos and the introduction of our domain expert. However, participants noted the lack of the author names and contact information. Similar to newspaper articles, \textbf{a responsible person should be named paired with, e.g., an e-mail address}. Additional sources for the context-driven parts of the story would further increase the participant's trust. Therefore, \textbf{links to credible internet sources}, e.g., the website of the WHO, can be provided. While our application supports multiple languages to be accessible for a greater audience, as an author one must deal with the problem that many external websites will not provide the information in multiple languages. \textbf{Additional information, e.g., metadata of a study, would also strengthen the trust of users} in the data-driven parts of our story. There is other work showing that the inclusion of audio output can influence the attitude towards the message~\cite{Meppelink2015}. Therefore, the inclusion of voice overs in the form of either a computer-generated voice, or voice recordings of the medical expert, might have a positive effect on the perceived credibility of the information.
Apart from the content, design plays an equally important role. A well-designed story can strengthen the interest as well as the credibility of the presented information. Therefore, \textbf{it is also important to include design expertise into the project}.

In particular, interaction design plays an important role. Our participants stated that the transition between the main story branch and optional side branches were not always clear. They also noticed missing affordances that emphasize interactive and clickable elements, leading to the concern about missing out content. Therefore, \textbf{navigation should be adjusted for switching between story branches}. Additional methods to highlight interaction possibilities should also be added. Since the story involves many different types of interaction, \textbf{it is important to emphasize not only that a slide is interactive, but also what interactions are available}.

\subsection{Narrative Visualization to Communicate Neurological Diseases}
Further generalizations for neurological diseases can be derived from our example of CSVD and the subsequent evaluation.
Most importantly, the creation process of a narrative medical visualization is highly interdisciplinary and requires at least one domain expert. In medicine, where things are often not clear, there ideally is a small team of experts reflecting different perspectives. This applies even more when it comes to treatment, where patients often ask for a second opinion that differs in many cases. People are accustomed to receiving information on healthcare topics from experts such as physicians. Therefore, \textbf{the physician should be introduced to the user during the story including their affiliations}, e.g., showing that they are working at a hospital or research institution, that highlight their expertise.\newline 

The brain is an organ with a unique shape, which is known even to non-experts. Therefore, \textbf{showing the brain as a whole is an appropriate anatomical context for general audiences}. However, for most aspects in neurology, there are complex technical terms, that do not have translations or colloquial synonyms, e.g., brain areas and in our case the sub-types of the disease itself. Ways have to be found to communicate these aspects to the audience without causing confusion.  We paired acronyms of complex terms with icons. However, users find it unsatisfactory and prefer to know the full terms even if they are more complicated or in a foreign language like Latin. Therefore, \textbf{even complicated terms should be introduced with their full name} if they cannot be paraphrased.

 A story about a disease showing, e.g., risk factors, symptoms or treatment, is complex. However, even if the goal is to inform about risk factors only, there must be an introductory section about the underlying disease if it is not commonly known. To be able to adapt a story to the user's personal knowledge level, it can be split up in sub-stories, like we have done within our nested story structure. While not implemented in our current story, it would be possible to add an advanced navigation method through the story that enables users to choose which parts to skip. A user who already knows a disease would not have to look at the parts of the story that introduce the disease. However, prevention possibilities could still be interesting for this person. Therefore, it is necessary that the sub-stories become more modular and work as independent stories while still fitting in the bigger scope of the overall story line, which we called a nested narrative structure. In this case, the author has to carefully consider the narrative intent of the overall story and decide which story parts are mandatory to communicate it. 

 During our evaluation, participants did not state that the story was too long. However, depending on where the story should be published, the time users want to invest in it may be very limited. Outside of an evaluation scenario, such educational stories could be shared via social media. People that accidentally come across such a story might have a much shorter time period available to view it as well as a shorter attention span. Therefore, investigating how narrative visualizations can be designed in a way such that they are adaptable to different audience groups with variable viewing times is important.\newline
 
\textbf{General explanations, e.g., about MRI, are reusable for other projects} since they have no contextual reference to the current story, but are useful background information for many different topics. MRI is particularly important for the diagnosis of many neurological diseases, such as multiple sclerosis and brain tumors, where the high soft-tissue contrast of this modality is beneficial.
 
In summary, we identified key aspects for creating narrative visualizations to communicate neurological diseases that are accepted by a general audience:
\begin{itemize}
    \item \textbf{A clear key message:} In disease stories, the key message conveys a certain call-to-action, e.g., to take preventive measures. It is strongly linked to the risk perception of the users, thus a clear key message is essential to convey and motivate preventive actions.  
    \item \textbf{A carefully thought-out story line:} The narrative must be built around the key message to support it in the most effective way. Information not necessary to convey it, might be placed in optional side branches. However, it is important that the side branches integrate well in the narrative flow to prevent users from missing the key message of the story.
    \item \textbf{Appealing design:} An appealing design supports the trust the users have in the presented information. Furthermore, a careful use of accent colors, icons, and pre-possessed as well as annotated medical data with appropriate anatomical context support user understanding of complex medical topics.
    \item \textbf{Easy-to-use interactions:} Medical experts navigate through medical volume data using interactions that a wider audience is not familiar with. Thus, it is important to provide simplified interactions for medical narrative visualizations.
    \item \textbf{Credible references:} To emphasize trust in the information presented is important for the users to accept the message. To create trust, it is important to show logos of involved institutions, medical experts as well as contactable authors of the story. The origin of context-driven content should be presented, e.g., by linking to external websites. For data-driven content, metadata about the study should be presented.
\end{itemize}

%

\section{Conclusion and Future Work}
Narrative visualization is an established field but previously not been applied to stories about medical diagnosis and treatment. Since medical data have unique characteristics, new solutions have to be found, e.g., for presenting 3D medical data. We presented a story for communicating the cerebral small vessel disease. We discuss how practices from narrative visualization can be adjusted for informing about neurological diseases. Based on our experience we propose a workflow for creating similar visual data stories as well as a structure for a patient-driven disease story. We also identify key aspects of designing such a disease story during our evaluation. 
Furthermore, our work contributes to the scientific outreach done by many health organizations, bridging the gap between healthcare communication and narrative visualization.

There are many possibilities for future research. Based on our current work, more flexible and interactive story structures can be explored. Furthermore, different genres to communicate medical topics can be used in future applications and evaluated regarding how they are perceived by different target audiences. Additionally, different narrative patterns need to be evaluated based on medical data by generating different versions of a story.
We designed a narrative visualization targeted at people that are not affected by CSVD. Our aim is to motivate them to minimize their risk factors. However, further research is needed in terms of accurate risk communication and perception. Additional feedback from psychologists would also be interesting as to what kind of information leads to a lasting change in behavior.

Narrative visualization might also target \emph{patient education}, e.g., explaining imaging methods, surgical interventions, and different treatment options. The logical next step would be to investigate how the explanation of a disease needs to be adapted when it is directed at patients instead of healthy individuals.\newline

\acknowledgments{
The authors wish to thank Benjamin Bach, University of Edinburgh for fruitful discussions of design decisions. 
}

\bibliographystyle{abbrv-doi}

\bibliography{template}
\end{document}